\documentclass[12pt]{amsart}

\usepackage{graphicx, verbatim,amsmath,amssymb, float}
\usepackage{hyperref}
\hypersetup{
    colorlinks=true,
    linkcolor=blue,
    filecolor=magenta,      
    urlcolor=cyan,
}

\def\vs#1 {\vskip#1truein}
\def\hs#1 {\hskip#1truein}
\def\eps{{\varepsilon}}

\newcommand{\nd}{\noindent}
 
\newcommand{\real}{{\rm I\kern-.2em R}}

\newcommand{\be}{\begin{equation}}
\newcommand{\ee}{\end{equation}}

\newtheorem*{thm*}{Theorem}

\everymath{\displaystyle}

\theoremstyle{definition}

\begin{document}

\title[Superfluid Helium]{Superfluid Helium}

\date{\today}

\author{Charles Radin}
\address{Charles Radin\\Department of Mathematics\\The University of
Texas at Austin\\ Austin, TX 78712} \email{radin@math.utexas.edu}


\begin{abstract}

  We present a {mathematically solvable toy model} of the phase
  transition between the normal { liquid} and superfluid phases of
  helium--4, and use it to show how a symmetry-based order parameter
  can be obtained for the superfluid. We also present a range of
  models in which one can obtain sharp bounds on the transition
  between the normal liquid and gas phases, arbitrarily close to
  the superfluid.
  \end{abstract}

\maketitle

\begin{figure}[H]
\center{\includegraphics[width=3in]{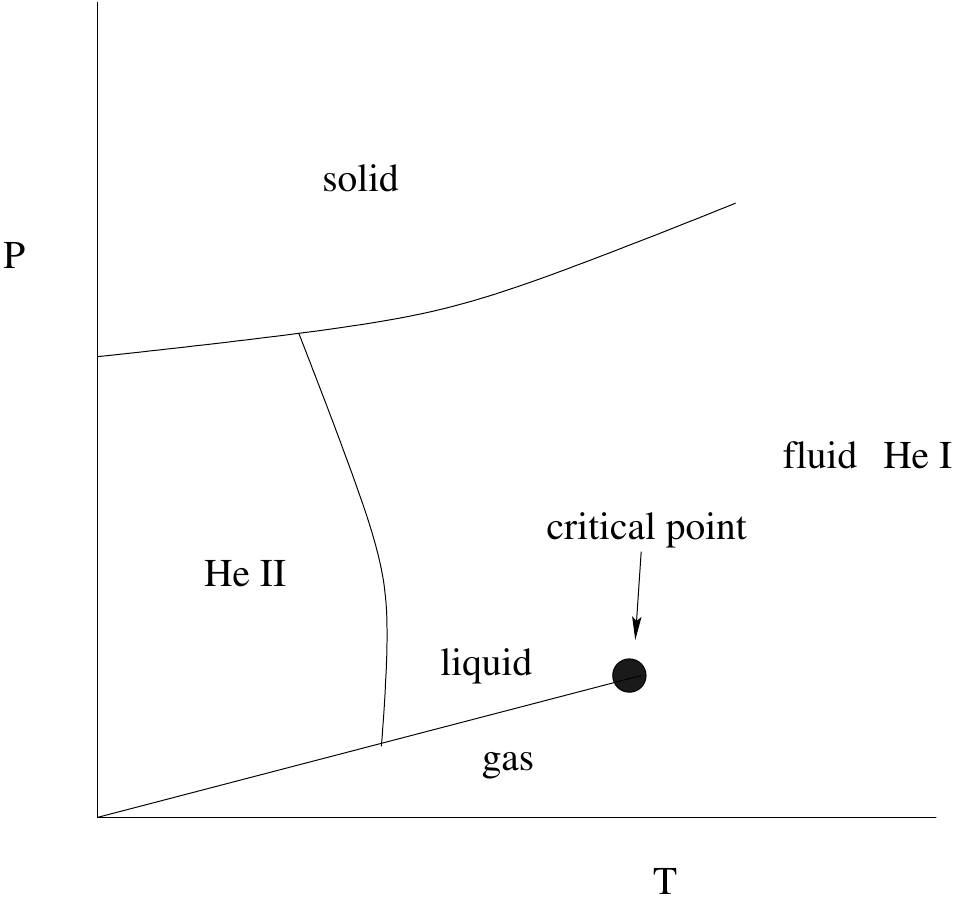}}
\vs.1 \caption{Highly simplified Pressure-Temperature phase diagram of $^4$He, showing
  the normal fluid (He I), the superfluid (He II), and a
  critical point within the normal fluid but not within the
  superfluid.}
\label{phases}
\end{figure}

Helium gas was liquified by Onnes in 1908 \cite{Onnes}, and the superfluid
form of the liquid was discovered in 1938 by Kapitza \cite{Kap}. In 1940 London
\cite{Lond} `associated' the superfluid with the dilute quantum gas
conjectured by Einstein in 1924-25 \cite{Ein}, with both the liquid and
dilute gas of $^4$He understood in
terms of Bose-Einstein statistics.  We present a toy model of $^4$He,
to play the role for superfluid $^4$He that is played by the Ising
models for fluids which exhibit a critical point. The physics is
fundamentally different for superfluid $^4$He, which does not exhibit
a critical point; see Figure \ref{phases}. Our model gives a
natural order parameter distinguishing the superfluid from the normal
liquid form of $^4$He, and also provides quantitative information
about singular behavior between the normal liquid and gas, all the way
to the point where that normal gas/liquid condensation meets the superfluid (see Figure \ref{phases}),
providing an understanding, in terms of Bose-Einstein statistics, of the role of emergent quantum behavior
near that point.

\begin{figure}[H]
\center{\includegraphics[width=4in]{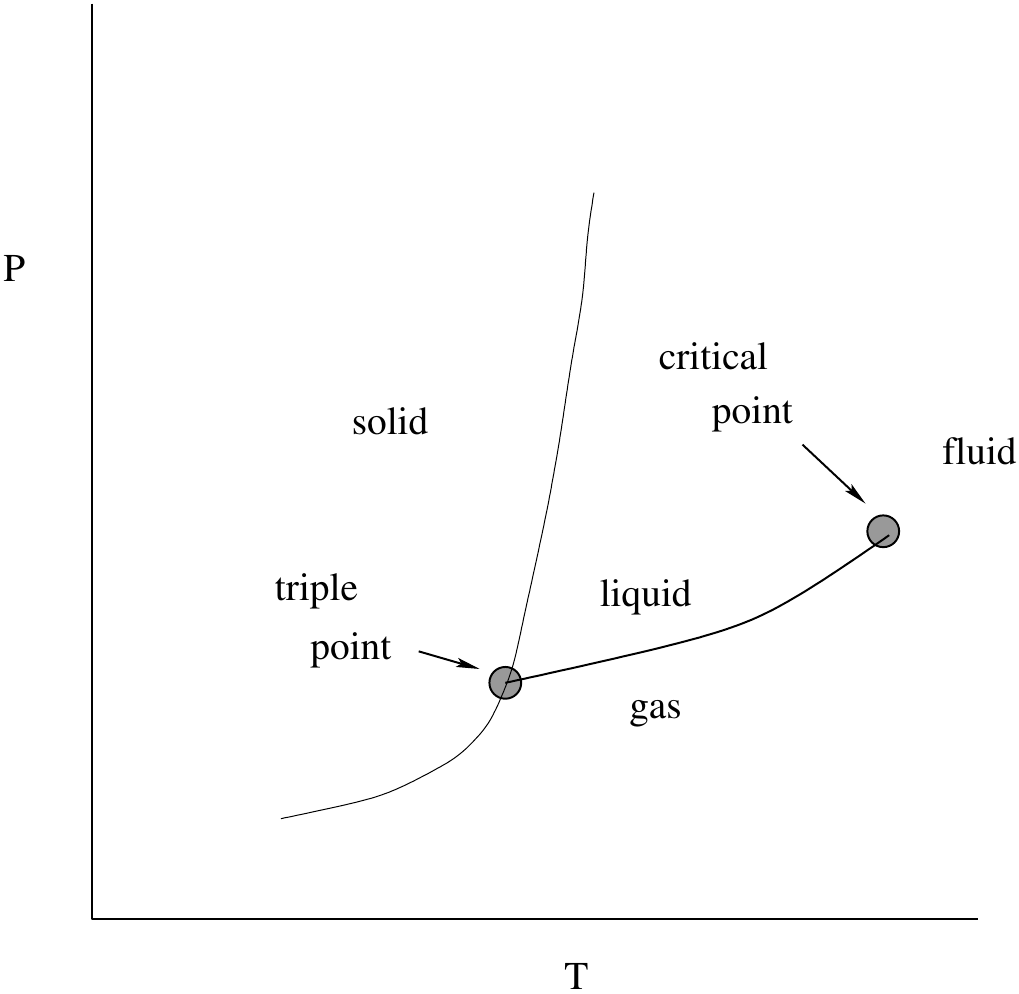}}
\vs.1 \caption{Highly simplified Pressure-Temperature phase diagram of carbon dioxide, emphasizing
  gas-liquid coexistence between the triple point and critical
  point. It also shows the solid/fluid transition, represented by the other curve
  going through the triple point.}
\label{carbon-dioxide}
\end{figure}

\begin{figure}[H]
\center{\includegraphics[width=3in]{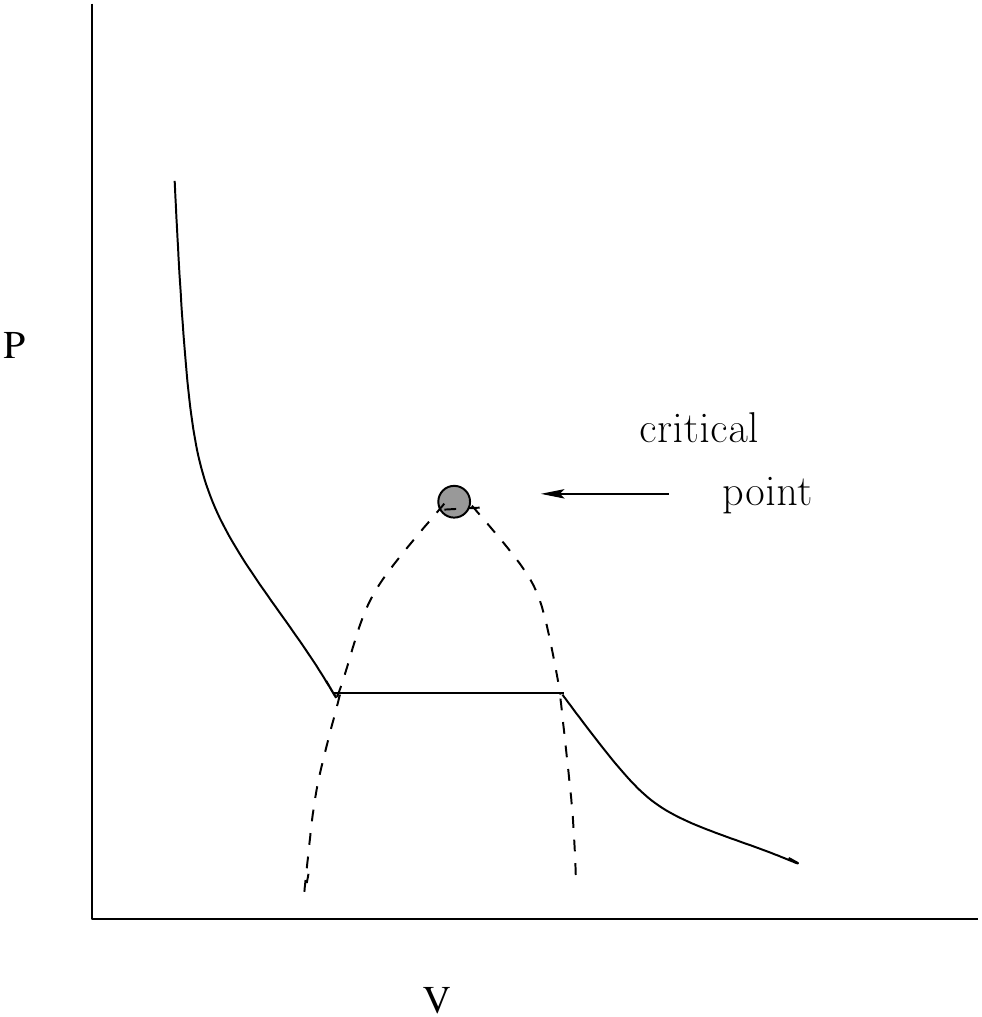}}
\vs.1 \caption{One isotherm (dark curve) of the highly simplified
  Pressure-Volume-Temperature  equation of state of carbon
  dioxide, the range of coexistence
indicated by the dashed curve.}
\label{isotherm}
\end{figure}

\vs.15 Motivated by the experimental condensation of carbon
dioxide gas by Andrews in the 1860's \cite{Andr}, in particular
the discovery of its critical point
(see Figure \ref{carbon-dioxide}), van der Waals
produced his
important model of gas/liquid condensation in 1873 \cite{VDW}, see Figure
\ref{isotherm}, significantly improved by Maxwell's `equal-area rule'
in 1875 \cite{Max1}.
\vs1
\vfill \eject

\nd (Van der Waals' equation of state incorrectly described
gas-liquid coexistence, which is correctly represented by the
horizontal line in a typical isotherm in Figure
\ref{isotherm}; Maxwell corrected this portion of the equation of
state with his `equal-area' rule.) The statistical mechanics of Boltzmann and
Maxwell \cite{Boltz, Max2} was still primitive at that time, but was put in mature form
by Gibbs {\it et al}$\,$ by around 1900 \cite{Gibbs}, and then was replaced by quantum
statistical mechanics when classical mechanics was replaced by quantum
mechanics, at the length scale of molecules, in the following decades
\cite{Dirac,vNeum}.
While chemistry uses statistical mechanics to study small
systems of particles, an important role of statistical mechanics is to
model the phases and phase transitions, appearing in the
thermodynamics of macroscopic matter in thermal equilibrium, as
originating in the mechanics of molecules \cite{Uhl1,Schr}.  The
formulae of equilibrium averages in (classical or quantum) statistical
mechanics, applied to a macroscopic but finite system of `particles',
are manifestly smooth as temperature and pressure vary, without the
sharp transition one would like to model \cite{KahUhl}.  
A careful analysis of the van der Waals model by Uhlenbeck, Kac,
Onsager, Penrose and others (see the history in \cite{deBoer, Ziff}), was
aimed at replacing Maxwell's ad hoc rule by an `infinite system limit'
formalism, applicable to a broader range of phenomena. This analysis
gradually led, between 1935 and 1970, to a class of useful models, not
just for the gas/liquid transition but also other transitions
with a critical point
such as the paramagnetic/ferromagnetic transition, the
order/disorder transitions of alloys, {\it etc}, notably including
the Ising models, in all of which perfectly sharp
transitions can be {\it proven to emerge within that framework} as the
system size grows unboundedly. (See \cite{Thomp} for a good presentation
of
these and other models.)
 The above is a brief overview of some
history relevant to this paper. A main point of this paper is that the
development of an infinite system limit between 1935 and 1970 is a
departure from the experimental phenomena associated with real finite
systems, and is worth analyzing to produce a useful model of
superfluids.
We note next some prominent omissions
in the above history, some of which offer important context for our
model of superfluid helium.

\vs.15 Although historically significant for a theory of
superfluidity we ignored Einstein's important contributions to
statistical mechanics. In particular we ignored his work on
Bose-Einstein Condensation \cite{Ein,Lieb}, as being too far
from our interest in superfluid He II, which is a dense liquid 
\cite{Lond}. This work of Einstein directly precipitated concern about
infinite system limits \cite{KahUhl, Ziff}. 
The
theory of the renormalization group \cite{Golden} seems relevant
but it is too specialized for our purposes and will not be
discussed. And finally, Bridgman's experiments around 1910 concerned
the distinction between fluids and solids in thermal equilbrium, which
led to an independent effort in developing infinite system limits in the
1950's to which we devote a separate discussion in the appendix, with
references, because of subtle points it raises about infinite system
limits.

\vs.15 Equilibrium statistical mechanics is a model of finite
systems of particles, modelling different forms of energy of the
systems, including heat, and was built with a variety of `ensembles',
each tailored to a certain way in which the system being modelled can
exchange energy with its environment. The
pressure-temperature ensemble assumes the system is in equilibrium in
an environment providing fixed temperature and pressure; the
microcanonical ensemble is for a system which is closed to all forms
of energy sharing with its environment  \cite{Gibbs,Salin}. One of the features of the
infinite system limit formalism developed in 1935--1970 was that
different ensembles give the same macroscopic thermodyamics
\cite{Rue},
though this was a convenient choice, not a necessity, and in
modelling superfluidity we will consider how such a
choice enters.  In this regard consider the Ising model, which was invented
to model the transition from paramagnetism to ferromagnetism, which
has a
critical point but which takes
place in certain solids such as nickel and iron \cite{Stef,Getz}, which is a serious
complication. To produce a {\it mathematically solvable} model of the
paramagnetic state, whereby a fluid not only freezes to a solid but
simultaneously does or does not produce magnetically ordered domains,
depending on temperature and pressure, one runs into the fundamental
difficulty that there has never been a mathematically solvable
statistical mechanics model of the freezing of liquid to solid, which
is part of this transition and {a major unsolved problem by
  itself}. (This is discussed further in the appendix.)
We emphasize that the Ising models do not attempt to model
the full phenomenon; they only model the emergence of a single
ordered domain within an {\it assumed} crystalline background. This
is clearly a subtle issue in the treatment of scaling of the system
size, meant to understand the emergence of ferromagnetism. We put off
to the appendix
further discussion of the fluid/solid
(melting) transition,
to discuss our approach to modelling superfluid helium.

\vs.15 The condensation transition is special in having a critical
point but dealing with the gas and liquid states of a fluid, as in
Figure \ref{carbon-dioxide}.  A significant feature of the gas/liquid
critical point in general, for instance in fluid He I which is
of particular interest to us is the fact that by adiabatically
varying temperature and pressure around it one may smoothly
transition from the gas to liquid states, which shows that those gas
and liquid states are, {\it by definition}, part of the same phase.
Experimentally there is no hint of a critical point
associated with the He I/He II transition (see Figure \ref{phases}), which raises the question
of how to model the transition.  For the condensation transition
between the gas and liquid parts of a single fluid phase (Figure
\ref{carbon-dioxide}) an infinite system limit formalism was created
and tailored to the purpose, as discussed briefly above.  That
formalism is not necessarily appropriate to analyze phase transitions
without a critical point,  such as the normal
liquid/solid, or He I/He II transition.  In this paper we develop an
appropriate infinite system limit formalism for the He I/He II
transition and model the transition {\it with intrinsically distinct phases},
distinguishing the phases by an order parameter.  We do this
indirectly in a toy model. Our argument is not appropriate for the
liquid/solid transition
and we will
see, along the way, `why' crystallinity has {\it not} successfully
provided an order parameter for the liquid/solid
(melting) transition. (See \cite{Ander} for a profound analysis of
phase transitions as
of 1983.)
We emphasize that the point of this paper is not to make a
model of a superfluid, like that of Landau \cite{LanLif} to be analyzed for
instance by simulation, but to make a toy model within which we can
analytically {\it prove} the emergence, and qualitative aspects, of features such as
the thermal conductivity or viscosity of $^4$He in {\it intrinsically
quantum terms.}


\vs.15 In modelling systems of ever larger size one needs to
normalize various quantities, for instance the energies in the toy model (playing the
role of kinetic energy and potential energy of interaction), and the
entropy.
Phase transitions are then signaled by nonanalyticity of the
entropy as a function of the normalized energies, in the infinite system
limit. We have to treat numerous limits carefully in their relation to the
infinite system limit, besides those mentioned above, all of which can
be explained most clearly in the context of a specific toy model.

\vs.15
The toy model of this paper is a graph model,
superficially similar but fundamentally different from that of Ising. In
their important paper from 1952 \cite{LeeYang}, Lee and Yang made a
lattice gas model out of the Ising model, and {\it using the infinite
  system formalism} of short range interactions those two models are
effectively interchangable \cite{Ginib}. But for a graph model with
long range interactions, such as ours, it becomes quite difficult to take an infinite graph
limit, comparable to the difficulty to obtain mathematical control of
the Ising model, even after many simplifications of the original proof of
Onsager from 1944 \cite{Onsag}. This is what has been done in the
model of this paper.
In this paper we slight the mathematics and concentrate on the
physics of the model, but it is appropriate to give some background on
the model. The model is set in the formalism of Lov\'{a}sz {\it et
  al} \cite{Lov}, which represents infinite graphs by
`graphons', functions $g(x,y)$ of $(x,y)$ in the unit square
$[0,1]\times [0,1]$ with certain properties. Our normalization of the
entropy has various origins, one of which is the fundamental large
deviations result of
Chatterjee/Varadhan \cite{CV}. The model is analogous to a
microcanonical ensemble but with a long range interaction, in essence
a mean-field model \cite{RadSad1}. For an expository presentation of
our model
including motivation and proofs see \cite{RadSadBook}.

\vs.15
In more detail, consider the set $S$ of all simple graphs with nodes on the finite
square grid $N\times N=\{(j,k): 1\le j\le N, \ \ 1\le k\le N\}$, with
undirected edges. If we associate particles with nodes, not allowing
multiple occupancy, then the maximum possible number of particles is
$N^2$.  Such graphs have at most $N(N-1)/2$ edges, so
if alternatively we associate particles with edges, not allowing
multiple occupancy, again there can be at most
order--$N^2$ particles. Both are well known ways of producing arrangements
of order--$N^2$ particles on the square grid, associating them with edges or
with nodes.
\vs.15
When particles are associated with nodes there is a clear connection
with the $2d$ Ising models, or the equivalent lattice gas version
noted above.
Alternatively, we illustrate the case where particles are associated with edges by
the edge-triangle model which is the basis of this paper. It uses two
energies. One is a chemical energy of value 1 associated with each
occupied edge.  The three-body triangle energy is the number of
occupied triangles; a triangle is a set of three edges corresponding
to nodes which could be labelled $(n_j, n_k)$, $(n_k,n_q)$, $(n_q,
n_j)$. The edge density is the number of occupied edges divided by
$N^2/2$, so the density is asymptotically in $[0,1]$, and the triangle
density is the number of occupied triangles divided by $N^3/6$,
so the density is again asymptotically in $[0,1]$.
\vs.15
The Ising or lattice gas model, and the edge-triangle model, become models
of statistical mechanics by use of a uniform probability distribution
on all particle arrangements with their two given energy
densities. Thus both models have two parameters, their two energy
densities, for simplicity denoted, only in this paragraph, by $d_1$ and $d_2$. The Boltzmann entropy
$B(d_1,d_2)$ is the natural logarithm of the number of particle
arrangements with densities $d_1$ and $d_2$, divided by $N^2$.  In
both models, with a finite number $N^2$ of particles the densities
have only a finite number of values but in an infinite system limit
those densities range over continuous values and for both models
singularities in the entropy $B(d_1,d_2)$ are proven to emerge as the
densities vary.

As noted above our edge-triangle model has two parameters: the edge
density $0\le \eps \le 1$ and the triangle density $0\le \tau \le
1$. The model is fully defined using the graph limit formalism and
then maximization of the Boltzmann entropy with constraints on the two
energy densities, a microcanonical ensemble. The
optimization is over graphons, using the LDP of
Chatterjee/Varadhan. This
gives the model a form of wave-particle duality familiar in quantum
mechanics, as follows.
\vs.15
Graphons represent the infinite graph limit by compressing the infinite
number of nodes on the $x$-axis into a dense
subset of the unit interval. To understand the wave-particle duality
one needs to be able to go back to finite systems, and show (see
section 2.6 in \cite{RadSadBook}, or \cite{NRadSad1}) how a
graphon, optimal for Boltzmann entropy constrained by edge and
triangle energy densities, implies well-defined probabilities for
other `macroscopic' observables, including the constrained Boltzmann
entropy, yielding a probabilistic interpretation, via the graphon
`field', of large systems of particles/edges.
The ultimate source for probabilities is in the
constrained Boltzmann entropy, the uniform (`microcanonical')
distribution over graphs with edge density $\eps$ and triangle density
$\tau$.

\begin{figure}[H]
\center{\includegraphics[width=4.5in]{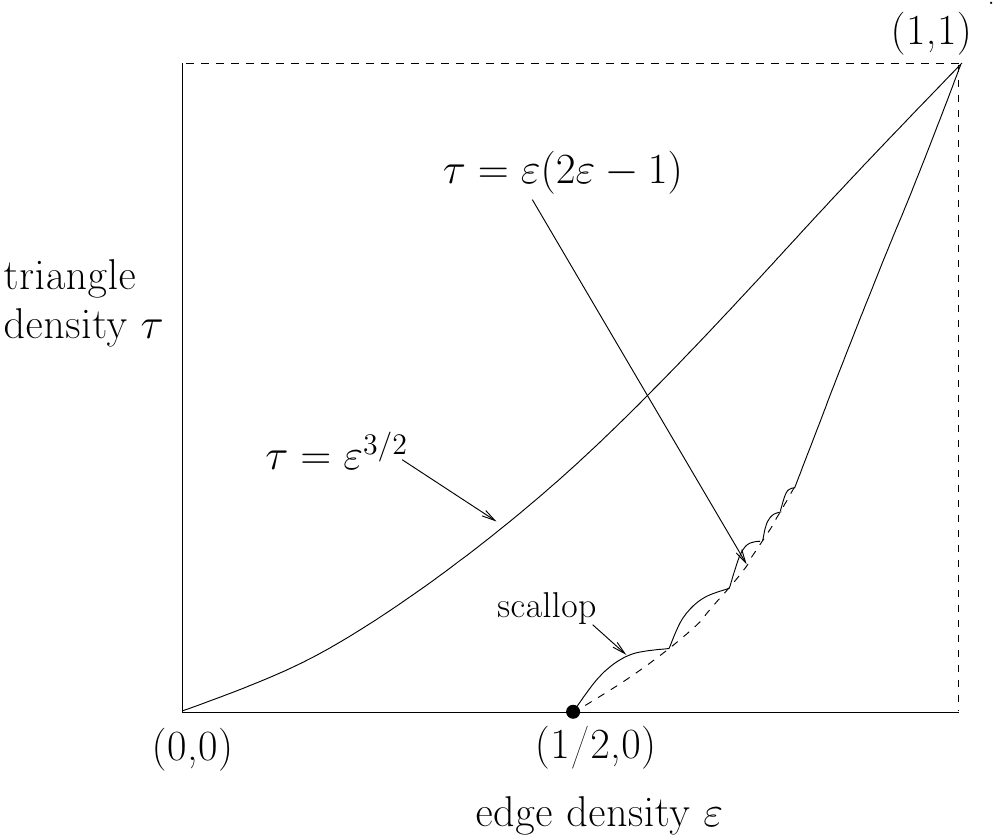}}
\vs.1 \caption{Razborov triangle, containing the possible values of edge
  and triangle densities.}
\label{Razborov}
\end{figure}

\vs.15
The `phase space' of the model is the set of limits, as $n\to \infty$, of
possible energy pairs $(\eps,\tau)$ and has been studied within
extremal combinatorics over many years \cite{PikRaz}, culminating in the boundary
and interior of the region shown in Figure \ref{Razborov}. (The rich
structure of the boundary of the phase space is another important support
for the above choices of normalization.) 
The phase space is bisected
into two subregions by the curve: $\tau=\eps^3$. On that curve the
equilibrium state, meaning the graphon optimizer of the entropy, is that of
{uniformly random edges for given constraint $\eps$}.

\begin{figure}[H]
\center{\includegraphics[width=5in]{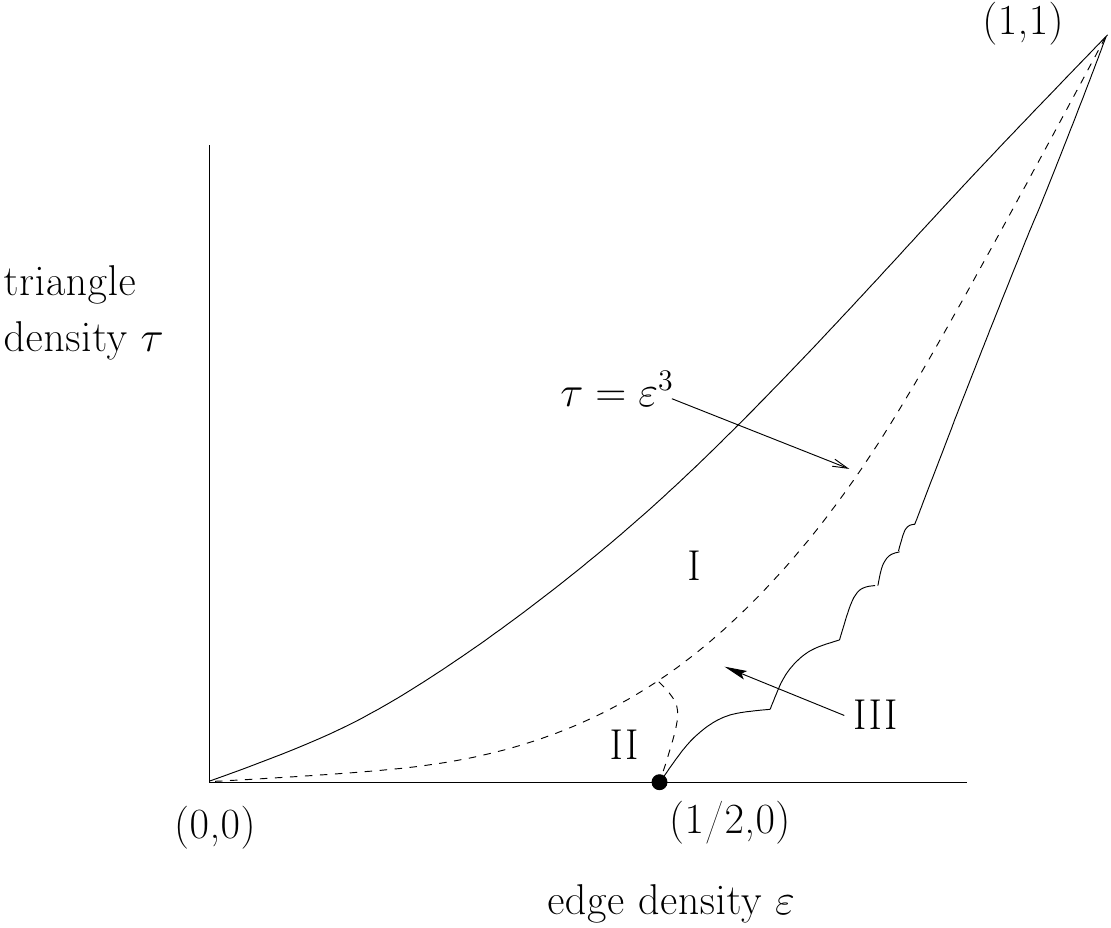}}
\vs.1 \caption{Some phases in the edge/triangle model}
\label{3phases}
\end{figure}

\begin{figure}[H]
\center{\includegraphics[width=3.5in]{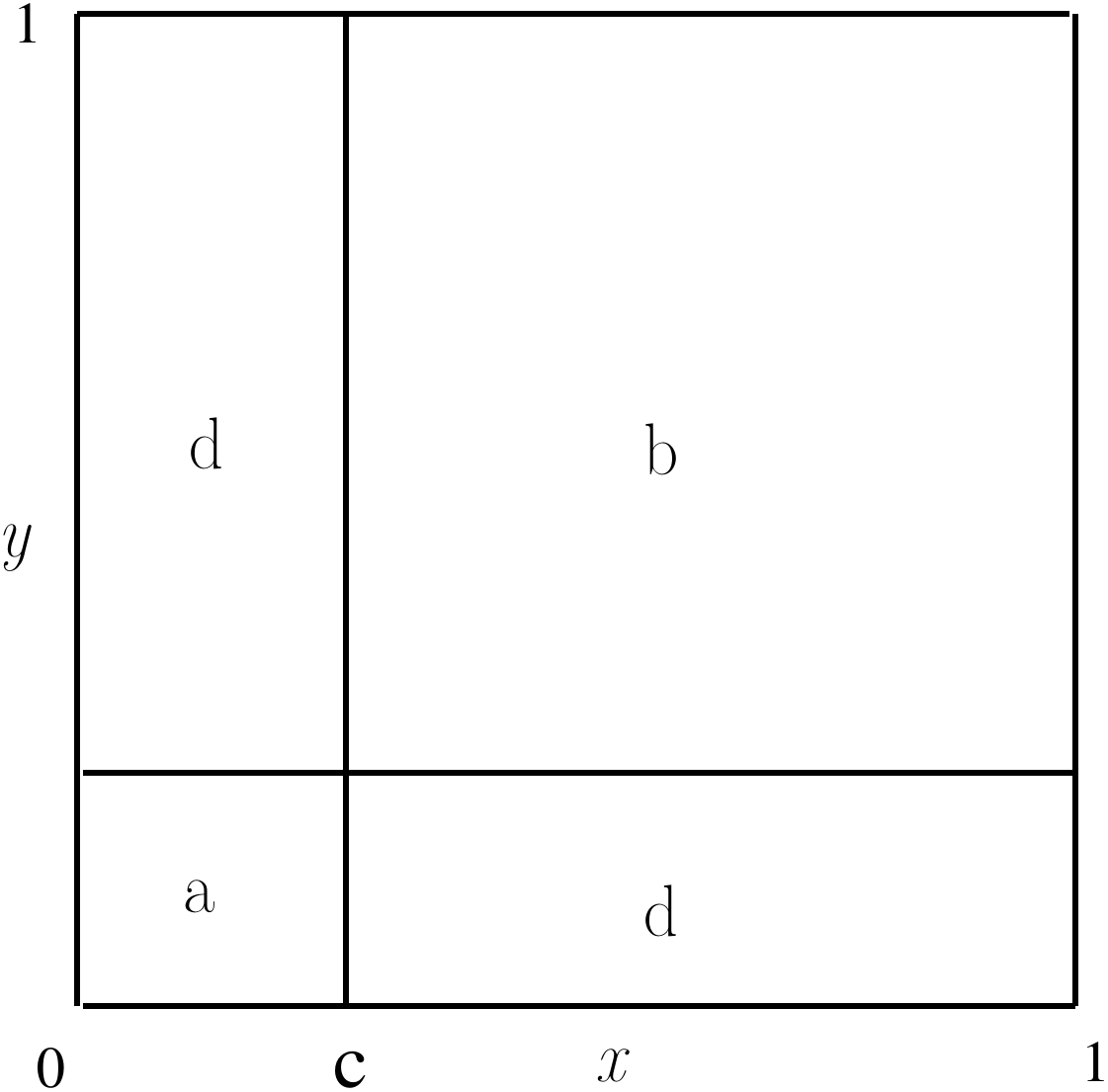}}
\vs.1 \caption{Structure of a bipodal graphon.}
\label{bipodal}
\end{figure}

\begin{figure}[H]
\center{\includegraphics[width=3.5in]{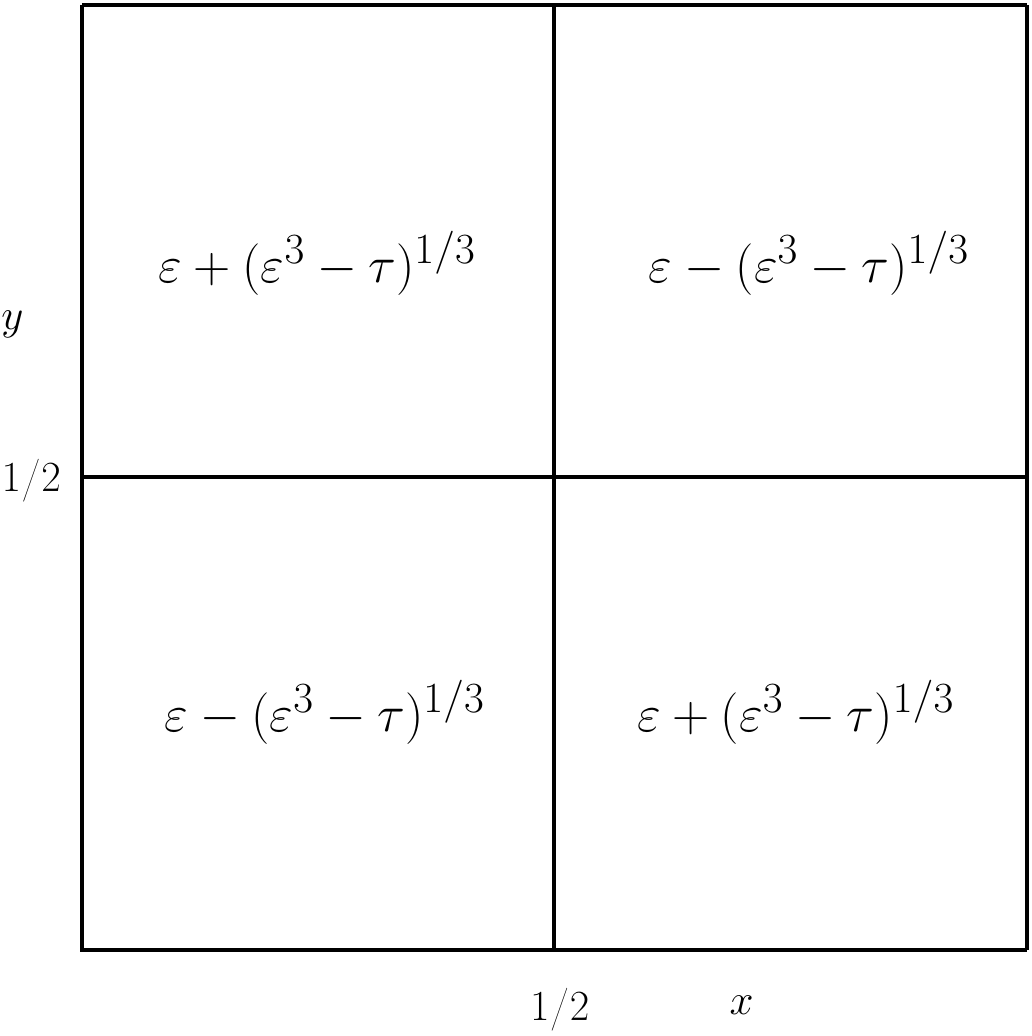}}
\vs.1 \caption{The symmetric A(2,0) graphon.}
\label{symmetric-bipodal}
\end{figure}

\vs.15
There are three subregions of the phase space of interest in this paper, open subsets
of the regions labelled I, II and III in Figure
\ref{3phases}. It has
been proven, in the infinite system formalism of the model, that the open
sets are each part of some phase \cite{RadSadBook,NRadSad1,KRRS,NRadSad2}. Graphs in these
phases are proven to be `bipodal', represented by
piecewise-constant graphons of the general form shown in Figure
\ref{bipodal}.  In this toy model I plays the role of the helium
gas, III the role of He I, and II the role of He II.  It is
proven \cite{RadSadBook,NRadSad1} that for $(\eps,\tau)$ in the region II the
optimizing graphon has the symmetric form of Figure
\ref{symmetric-bipodal}.

\vs.15 It has also been proven in \cite{RadSadBook,NRadSad1} that the difference between the
nonsymmetric bipodal graphon representing $(\eps,\tau)$ in region III and
the symmetric graphon representing $(\eps,\tau)$ in region I cannot be
analytic continuations of one another, and therefore constitute an
order parameter distinguishing the two phases in an {\it essential}
way. Furthermore, the precise singularity of the entropy, and a range
of other observables, has been determined all along both sides of the
transition between regions I and III, representing the gas and normal
liquid. Although we have concentrated on the edge/triangle model,
in fact our techniques, and many proofs, in \cite{RadSadBook} apply
to a wide range of graph models, which might be of value in
studying the transition between the gas and normal liquid of $^4$He.

In summary, the mathematical model in this paper is an
application of the infinite system limit for the mathematical system
of dense graphs, starting from the fundamental results of Lov\'{a}sz
{\it et al} \cite{Lov}, Chatterjee/Varadhan \cite{CV}, and extreme
combinatorics \cite{PikRaz}, adding the use of an analogue of the
entropy of Boltzmann \cite{RadSadBook,RadSad1,RadSad3}.  By necessity we consider
many limits in our toy model of superfluid helium.  On the
experimental side we use a thermal equilibrium or infinite time limit,
and we also employ a lattice model which removes the continuum limit
common in the description of macroscopic matter.  Since the goal is to
understand large but finite constrained graphs, over finite time, we have chosen how to
normalize relevant quantities and take limits so as to obtain useful
information about finite samples of helium, namely the structure of
phases which emerge with increasing system size.  One cannot prove
that this way to take limits is necessary. Rather, we have chosen this
path to show two things: how a
symmetry-based order parameter appears naturally for superfluid
$^4$He, and that some of the singularities of the gas/liquid transitions of $^4$He can be computed. It is
also perhaps a welcome feature of the model that the quantum mechanical nature of the
transition, at low temperature and pressure, commonly associated, since London \cite{Lond}, with either
Bose-Einstein statistics in an urn model for condensates at low
density, or bosonic exchange symmetry of the wave function, is captured
in the graphon model by an analogue of wave-particle duality. In
the appendix we have contrasted our approach with the still
open problem of obtaining an order parameter for the more famous
fluid/solid transition.

\vfill \eject


\vfill \eject
\hbox{}
\vs.2
\centerline{{\bf Appendix on the fluid/solid transition}}
\vs.2
We now return to the physics,
to discuss a {\it secondary goal} of the paper,
concerning the fluid/solid
transition.
Starting from his PhD thesis of 1908 Bridgman developed
experimental high-pressure techniques which gave convincing evidence
-- he received the Nobel prize in physics for this in 1946  --
that the fluid and solid phases of matter are intrinsically different:
they cannot be adiabatically transformed into one another, at least at
high pressure and high temperature. (The fact that they also
cannot be connected smoothly at low temperature was determined, in the
following decades, to be a quantum mechanical effect, and a secondary
goal of this paper is to clarify, in a small way, the theoretical aspects of this
low temperature portion of the Bridgman result. The analogue of
wave-particle duality in the model is suggestive.) 
Bridgman's  high pressure phenomenon had a big impact on
equilibrium statistical mechanics, of which we concentrate on two threads.
In 1953--1955, Rosenbluth, Alder, Wainwright, Wood {\it et al}
(see
\cite{Wood} for a good history) invented algorithms to use on the
programable computer MANIAC at Los Alamos, on systems of 2-dimensional
`hard disks' and of 3-dimensional `hard spheres' \cite{Lowen}. In these equilibrium
statistical mechanics models extended particles are assumed not to be deformable or
lose energy to friction or viscosity. Their
results eventually gave convincing evidence of a transition in these
models by about 1963 \cite{Temp}, the `Kirkwood transition'
predicted in 1948 by Kirkwood \cite{Kirk}.
(We will discuss at various places below the strong assumptions on energy sharing in
the model, as well as the continuous positional range of the particles.)
One check on the theory of the model came in
physical `granular' experiments begun by Bernal \cite{Bern} and Scott
\cite{ScottGD} in
1959--1960, referring to the Los Alamos simulations. This
eventually led, starting in 1976, to a series of physical cyclic
shear experiments \cite{ScottAM, Nicol, Rietz}{\it etc}, 
using relatively hard spherical particles 
of approximately uniform radius and,
more convincingly, relatively hard disks \cite{Su} of approximately uniform
radius, in the very different scales of thousands (millions in 3d) of
macroscopic particles/granules of roughly millimeter size. Both the disks and
spheres were subjected to cyclic shear under controlled cycle
frequency and amplitude, and external
pressure, and as macroscopic particles they can be analyzed using
Newtonian mechanics \cite{Jin}, taking into account loss of energy to
friction, viscosity and deformation. They show an order/disorder
transition,
but more specifically these experiments confirm
the important role played by orientational order, rather than
crystallinity, in the hard sphere models \cite{Su}. However we emphasize
that the cyclic shear experiments involve very different physics; they
cannot be treated as merely a change of scale of $^4$He
atoms. (In terms of length scale there is an important
intermediate category between molecules and granular matter, called
colloids. For our purposes colloids are fundamentally less interesting
because of
the significant contribution of gravity to the energy of the system; 
see section 3.2 in \cite{RaSw}.) However both the theory of the hard sphere model
and experimental results on cyclically sheared macroscopic granules
can be treated by
the basic method of equilibrium statistical mechanics, keeping track
of conversions of energy \cite{Engel,Jin}. Given the prominence of
the hard sphere model in condensed matter theory over many years \cite{Lowen}
it is significant that there is no proof of a phase transition in the
model,
which
we take as evidence of an essential weakness as a model
for the fluid/solid transition.

\vs.1 
\begin{figure}[H]
\center{\includegraphics[width=4in]{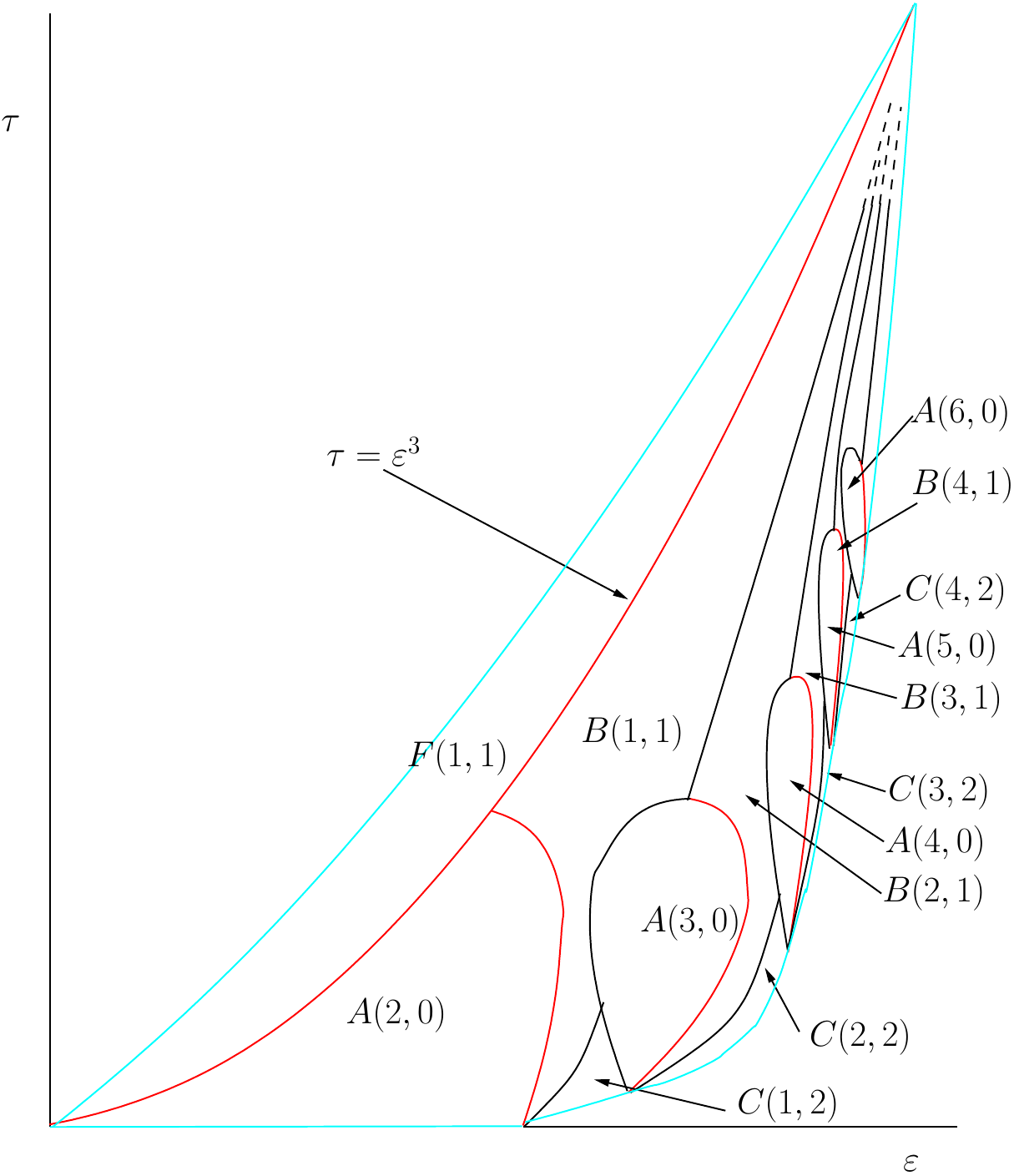}}
\vs.1 \caption{Conjectured phases (2017) in the edge/triangle model}
\label{conjectured-phases}
\end{figure}

At this point we contrast the transition between the
liquids He I and He II of Figure \ref{phases} and the melting
transition between solid and fluid carbon dioxide of Figure
\ref{carbon-dioxide}, both discussed above. In our toy model of $^4$He
we use a difference in symmetry to model an order parameter. To create
an order parameter for the melting transition one needs to understand
what is essentially different between solids and fluids in thermal
equilibrium.  There is a famous argument due to Landau (discussed in
\cite{Ander})
that
claims to distinguish fluid from solid by the difference in symmetry
of a homogeneous fluid and a crystalline solid.  However there is an
intrinsic difficulty in such a use of `crystallinity' since that is a
highly degenerate criterion, with unit cells of arbitrary size.  (Not
only does manganese have a large unit cell under common temperature
and pressure, there are alloys which are quasicrystalline,
which can be interpreted as having an infinite unit cell.)  Our edge/triangle model
suggests how to handle the degeneracy, using phases of our model
that we have not yet mentioned, labelled $C(1,2),
C(2,2), C(3,2) \cdots$  at the bottom of the
phase diagram in Figure \ref{conjectured-phases} (from \cite{Ken}). It has been proven
\cite{RadSadBook} that those phases are represented by graphons with more and more
complicated pode structures: 3-podal, 4-podal {\it etc}, of different symmetry, very much like
crystals with complicated unit cells; see Figures \ref{bipodal} and \ref{symmetric-bipodal}.
This suggests that to make use
of crystallinity as an order parameter for equilibrium solids one
might naturally use an infinite system limit as in the edge/triangle
model, appropriate to a model with long-range interaction.
We
emphasize that we are not claiming how to make an order parameter for
the fluid/solid transition (see however \cite{AriRad, Su}), an important open
problem \cite{Bru, Uhl1}; our arguments are only directly applicable
to the  transition between the normal fluid and superfluid phases of
$^4$He.

\vfill \eject

  \nd

\end{document}